\documentclass[aip,jcp,amsmath,amssymb,reprint]{revtex4-2}
\usepackage[dvipdfmx]{graphicx,color}
\usepackage{here}
\usepackage{bmpsize}
\usepackage{epsf}
\usepackage{bm}
\usepackage[usenames,dvipsnames]{xcolor}
\usepackage[normalem]{ulem}
\usepackage{txfonts}
\usepackage{lipsum}

\begin{document}

\title{
Simulating the nematic-isotropic phase transition of liquid crystal
model via generalized replica-exchange method
}

\author{Kengo Takemoto}
\affiliation{
Division of Chemical Engineering,
Graduate School of Engineering Science, Osaka University, Toyonaka, Osaka 560-8531, Japan
}

\author{Yoshiki Ishii}
\affiliation{Graduate School of Information Science, University of Hyogo, Kobe, Hyogo 650-0047, Japan}

\author{Hitoshi Washizu}
\affiliation{Graduate School of Information Science, University of Hyogo, Kobe, Hyogo 650-0047, Japan}

\author{Kang Kim}
\email{kk@cheng.es.osaka-u.ac.jp}
\affiliation{
Division of Chemical Engineering,
Graduate School of Engineering Science, Osaka University, Toyonaka, Osaka 560-8531, Japan
}

\author{Nobuyuki Matubayasi}
\email{nobuyuki@cheng.es.osaka-u.ac.jp}
\affiliation{
Division of Chemical Engineering,
Graduate School of Engineering Science, Osaka University, Toyonaka, Osaka 560-8531, Japan
}

\date{\today}

\begin{abstract}
The nematic-isotropic (NI) phase transition of 
4-cyano-4'-pentylbiphenyl (5CB) was
 simulated 
using the generalized replica-exchange method (gREM) based on molecular
 dynamics simulations.
 The effective temperature is introduced in gREM, allowing the
 enhanced sampling of 
 configurations in the unstable region, which is intrinsic to 
 the first-order phase transition.
The sampling performance was analyzed with different system sizes and
 compared with that of 
 the temperature replica-exchange method (tREM).
It was observed that gREM is capable of sampling configurations at
 sufficient 
 replica-exchange acceptance ratios even around the NI transition temperature.
A bimodal distribution of the order parameter at
 the transition region was found, which is in agreement with the
 mean-field theory.
In contrast, tREM is ineffective around 
 the transition temperature owing
 to the potential energy gap between the nematic and isotropic phases.
\end{abstract}

\maketitle

\section{Introduction}

Highly anisotropic molecules can form a variety of 
mesophases, including nematic, smectic, and columnar phases
between crystalline solids and isotropic liquids.~\cite{deGennes:1995vq}
Such molecules exhibiting liquid crystal (LC) phases are referred to as
mesogens.
One of the common mesogens is 4-cyano-4'-pentylbiphenyl
(5CB), which undergoes a nematic-isotropic (NI) phase transition at 
room temperature.
An order parameter for characterizing the
nematic LC phase is introduced with respect to the director, which is
defined as a collection of mesogen molecules.
The order parameter 
reduces with an increase in temperature.
Typically, it reduces from 0.6 to 0.4 on 
approaching to the transition temperature
and exhibits a discontinuous drop to zero at the transition.
This behavior is regarded as a sign of the first-order phase
transition, as demonstrated well by the mean-field theory for the NI
transition.~\cite{Stephen:1974kb, Singh:2000cv, Andrienko:2018cs}

Because the time scale associated with the equilibration of the nematic phase
becomes longer with an isotropic starting configuration,
it is still challenging to perform molecular 
dynamics (MD) simulations of the NI phase transition from atomistic levels,
even though 
considerable research has been conducted on the use of all-atom, united-atom (UA),
and coarse-grained models.~\cite{Zannoni:2001hu, Care:2005ju,
Wilson:2005hu, Wilson:2007gk,
Berardi:2008be, Bag:2016wg, Zannoni:2018gx, Allen:2019dd}
In particular, 
MD simulations of systems exhibiting first-order phase
transitions often encounter difficulties in sampling the configurations
around the transition 
temperatures, owing to the presence of unstable states
bridging the potential energy gaps between two
stable phases.

The temperature replica-exchange method
(tREM) (also known as parallel tempering) is a promising sampling 
method for 
simulating the phase transitions in various physical and chemical
systems.~\cite{Hukushima:1996dn, Sugita:1999cl}
The tREM is developed to 
sample a wide range of configurations, where
many replicas at different temperatures are simulated
in parallel and the configurations between two replicas are exchanged
to prevent replicas at lower temperatures from staying in the 
local minimum state.
However, the 
exchanges of replicas become ineffective near the first-order
phase transition temperature, which results in tREM inefficiency.
It should be noted that various enhanced sampling methods beyond the
original tREM have 
been developed.~\cite{Okamoto:2004cb}
Berardi
\textit{et al.}~\cite{Berardi:2009fk} and Kowaguchi
\textit{et al.}~\cite{Kowaguchi:2021jy} attempted to improve the sampling efficiency near the NI
phase transition with 
the Gay--Berne model and anisotropic Lennard-Jones fluids using the
multidimensional
REM~\cite{Sugita:2000kk} (also
known as Hamiltonian REM~\cite{Fukunishi:2002bc})
 and the isobaric-isothermal
REM,~\cite{Okabe:2001dn} respectively.

Recently, 
Kim \textit{et al.} proposed
the generalized replica-exchange method (gREM),~\cite{Kim:2010kn} which
was designed to efficiently simulate first-order phase transition systems.
In gREM, the effective potential and effective temperature are
introduced for each replica to ensure the unimodal distribution of the
potential energy $E$ of the system, even in the thermodynamically unstable
region. 
The effective potential is derived from the ensemble weight, which is
set to the Tsallis form in practical implementations.
Further, by utilizing its dependence on the most probable value of $E$, the effective
temperature is connected to the statistical temperature of the system.
Sufficient overlaps were achieved for the probability
distributions of $E$ for replicas with different effective
temperatures, which enabled efficient sampling of the transition region.
The gREM has been applied to various systems showing 
solid-liquid~\cite{Lu:2012jr, Lu:2014cl} and
vapor-liquid~\cite{Lu:2013kv, Ballal:2019ij} transitions.

The isobaric extension of gREM was proposed by
Ma\l olepsza and Keyes.~\cite{Maiolepsza:2015jj, Maiolepsza:2015kc,
Maiolepsza:2015co, Maiolepsza:2015dn}
In this extension, the enthalpy $H$ is used instead of the potential
energy $E$, which is employed in the canonical case (to be exact, $H$ refers to the
sum of $E$ and the product of pressure and volume).
The effective temperature for
each replica is introduced with respect to $H$, and owing to the use of $H$,
gREM is appropriate for simulating constant-pressure ensembles.
The isobaric gREM was also utilized to investigate the liquid and vapor 
phases of water~\cite{Cho:2014hw} and the order-disorder phase transitions
in lipid bilayer systems.~\cite{Stelter:2017cw}

Therefore, it is of interest to assess the applicability of isobaric gREM
to the NI phase transition, which is characterized by the orientational ordering.
In this study, we performed MD simulations in combination with isobaric
gREM for 5CB based on the united-atom (UA) 
model developed by Tiberio \textit{et al}.~\cite{Tiberio:2009iq}
Finite-size effects on the NI 
transition are important because an equilibration time scale near the phase
transition region drastically increases with
increasing the number of molecules $N$.
In computer simulations for LC systems, the finite-size effects were indeed examined,
for example, using 
the Lebwohl--Lasher model~\cite{Zhang:1992ju, Fish:2009ek, Shekhar:2012gk} and
anisotropic Lennard-Jones fluids.~\cite{Greschek:2011fw}
We investigated system size effects
by the use of gREM with the number of molecules $N$ ranging from 250 to 4000.
The sampling performance was also analyzed by comparison with tREM.
From sampled configurations, we examine the temperature dependence of
the density and order parameter
and discuss the temperature and enthalpy relationship.

\section{Model and Simulation details}
\label{Sec:Methods}

The UA model for 5CB developed by Tiberio \textit{et al.} was utilized in
this study.
In this model, the generalized AMBER force field (GAFF) was modified 
to reproduce the temperature of the NI phase
transition in the experiments, $T_\mathrm{NI}\approx 308.2$ K.~\cite{Tiberio:2009iq}
The comparison with experimental
results of the density and order parameter was also reported in Ref.~\onlinecite{Tiberio:2009iq}.
This UA model has also been employed in previous studies,~\cite{Sidky:2018hr,
Shi:2020cb, Sheavly:2020bn} which considered
the NI phase transition in a consistent manner with
Ref.~\onlinecite{Tiberio:2009iq}.
Another UA model was developed by the parametrization of the
TraPPE-UA force field for the 5CB molecule.~\cite{Zhang:2011eq}
Moreover, 
the transferability of the coarse-grained model has also been
proposed.~\cite{Zhang:2012fw, Zhang:2014jc}
The present work is mainly methodological, and Tiberio \textit{et al.}'s
model was
employed since it reflects the chemical reality and reproduces the
NI transition temperature well.

The simulated system was composed of 5CB molecules in a
cubic box with periodic boundary conditions.
The number of molecules was varied as $N=250$, 1000, 2000, and 4000.
The pressure was set to 1 atm, and the temperatures were ranged between
300 K and 320 K.
The $NPT$ ensemble with 
the N\'{o}se--Hoover thermostat and isotropic Parrinello--Rahman barostat
was used with a time step of 2 fs in all the simulations.
We performed gREM and tREM using the Large-scale Atomic/Molecular
Massively Parallel Simulator (LAMMPS)~\cite{Plimpton:1995wl}.

\begin{table}[t]
\caption{\label{table:enthalpy}
Enthalpy at 300 K and 320K (denoted by $\tilde{H}_1$ and $\tilde{H}_M$ for
 the use of gREM, respectively),
 and slope $\gamma$ for the gREM at the examined
 system sizes.
The enthalpy is normalized by the number of molecules $N$.}
\begin{ruledtabular}
\begin{tabular}{ccccc}
 $N$  & 250 & 1000 &  2000  & 4000\\
\hline
 $\tilde{H}_1 / N$  (kcal$\cdot$mol$^{-1}$) & 69.4 & 69.4 &  69.5 & 69.4\\
 $\tilde{H}_M / N$  (kcal$\cdot$mol$^{-1}$) & 71.9 & 71.9 &  71.9 & 71.9\\
 $\gamma$  ($10^{-3}$ K/kcal$\cdot$mol$^{-1}$) & -31.9 & -8.14 & -4.11 & -2.04
\end{tabular}
\end{ruledtabular}
\end{table}

The basic idea of gREM is summarized as follows (see the detail in Ref.~\onlinecite{Kim:2010kn}):
The effective temperature $T_\alpha$ of replica $\alpha$  (=1, 2,
$\cdots$, $M$)
is introduced by an inverse mapping of the effective potential
$w_\alpha$, $T_{\alpha}=(\partial w_\alpha/\partial E)^{-1}$.
$w_\alpha$ is related to the ensemble weight $W_\alpha$ through
the relation, $w_\alpha=-\ln W_\alpha$.
$T_\alpha$ is set to intersect the statistical
temperature, $T_\mathrm{s}=(\partial S/\partial E)^{-1}$.
Here, $S$ and $E$ denote the configurational entropy and potential energy
in the canonical ensemble, respectively.
Furthermore, the parametrization with a linear function for the effective temperature,
$T_\alpha(E)=\lambda_\alpha+\gamma(E-E_0)$, provides 
the ensemble weight that is equivalent with the form of the Tsallis
statistics, $W_\alpha\sim [\lambda_\alpha+\gamma(E-E_0)]^{-1/\gamma}$.
Here, $E_0$ represents an arbitrarily chosen potential energy and
$\gamma$ is chosen to have a sufficiently
large negative value to ensure that $T_\alpha(E)$ intersects $T_s(E)$ only once.
Note that 
$\lambda_\alpha$ is a control parameter for
determining the distribution of $E$ in
replica $\alpha$ for a given slope $\gamma$.
The non-Boltzmann ensemble with the weight $W_\alpha$ and effective 
temperature $T_\alpha$ enables to sample the thermodynamically unstable
states bridging two stable phases.

The isobaric gREM is formalized considering the
enthalpy $H$, statistical
temperature $T_\mathrm{s}=(\partial S/\partial H)^{-1}$, and effective
temperature $T_\alpha=(\partial w_\alpha/\partial H)^{-1}$ for replica
$\alpha$.~\cite{Maiolepsza:2015jj}
In isobaric gREM, the enthalpy dependent
effective temperature for replica $\alpha$ is given by 
\begin{equation}
T_\mathrm{\alpha}(H)=\lambda_\alpha + \gamma (H-H_0),
\label{eq:Teff}
\end{equation}
where 
$H$ is the sum of the potential energy and the product of pressure
and volume, and 
$H_0$ and $\gamma$ represent the reference enthalpy and the slope of
$T_\alpha(H)$, respectively.
Again, $\gamma$ and $\lambda_\alpha$ are control parameters of the
temperature intercept at a chosen enthalpy $H_0$.
The acceptance probability of replica exchange between neighboring
replicas $\alpha$ (with enthalpy $H$) and $\alpha'$
(with enthalpy $H'$) is given by
\begin{equation}
A_{\alpha, \alpha'} = \min[1, \exp(\Delta_{\alpha,\alpha'})],
\label{eq:Metropolis}
\end{equation}
with $\Delta_{\alpha,\alpha'}=w_{\alpha'}(H')-w_{\alpha'}(H)+
w_\alpha(H) - w_\alpha(H')$.
In contrast, the acceptance of replica exchange in tREM is given
by Eq.~(\ref{eq:Metropolis}) with 
$\Delta_{\alpha,\alpha'}=
(\beta_\alpha-\beta_{\alpha'})
(E_\alpha-E_{\alpha'})$, where
$\beta_\alpha=1/T_\alpha$ and $\beta_{\alpha'}=1/T_{\alpha'}$ are inverse
temperatures of replica $\alpha$ and $\alpha'$, respectively.
The volume is equilibrated with the barostat of 1 atm if the replica
exchange is occurred.

In the present simulations, $M=11$ replicas were utilized for both gREM and tREM.
The lowest and highest temperatures were 300 K and 320 K; hence,
temperatures of replicas 1 and $M$ were $T_1=300$ K and $T_M=320$ K,
respectively.
The values of enthalpy at the two temperatures (denoted by $\tilde{H}_1$ and
$\tilde{H}_M$, respectively) were quantified from the
averages of 10 ns simulations after equilibration over 50 ns
using the \textit{NPT} ensemble.
For supplementary comparison, we also simulated isobaric-isothermal REM by Okabe \textit{et
al.}~\cite{Okabe:2001dn} with $N=4000$, where $\Delta_{\alpha,\alpha'}=(\beta_\alpha-\beta_{\alpha'})
(E_\alpha-E_{\alpha'})+(\beta_\alpha p_\alpha-
\beta_{\alpha'} p_{\alpha'})(V_\alpha-V_{\alpha'})$ is used in
Eq.~(\ref{eq:Metropolis}) with the pressure and volume pairs $(p_\alpha,
V_\alpha)$ and $(p_{\alpha'}, V_{\alpha}')$ of replica $\alpha$ and
$\alpha'$, respectively.
The second term is designed 
to employ the enthalpic quantity for generation of an isobaric-isothermal ensemble.

For gREM, the slope $\gamma$ was
determined from $\gamma=(T_M-T_1)/(\tilde{H}_1-\tilde{H}_M)$.
The values of $\tilde{H}_1$, 
$\tilde{H}_M$, and $\gamma$ are listed in Table~\ref{table:enthalpy}.
The reference enthalpy $H_0$ in Eq.~(\ref{eq:Teff}) was set to
$\tilde{H}_1$ in the present gREM.
It is also seen that $\tilde{H_1}/N$ and $\tilde{H_M}/N$ are insensitive to
the system size; therefore, $\gamma$ varies roughly in inverse proportion to
$N$. 
The control parameter $\lambda_\alpha$ for replica $\alpha$ is given by
$\lambda_\alpha=\lambda_1+\Delta \lambda(\alpha-1)$ with $\Delta
\lambda= (\lambda_M-\lambda_1)/(M-1)$, where 
$\lambda_1=T_1$ and $\lambda_M=T_M-\gamma(\tilde{H}_M-\tilde{H}_1)$.
In contrast, for tREM, the temperature of replica $\alpha$ is 
given by 
$T_\alpha=T_1 + \Delta T(\alpha-1)$ with $\Delta T=2$ K.
A 50 ns simulation was run, in which a temperature swap between adjacent
replicas were attempted
every 10 ps both for both gREM and tREM.
We calculated various quantities presented in Sec.~\ref{sec:results} from
the last 10 ns simulation containing 1000 configurations, where the
stationary state was achieved.
Since all the events of temperature swap were tracked, the data can be
extracted for each replica in both gREM and tREM.

Note that the most probable enthalpy $H^*$ in gREM provides the 
relation, 
$T_\mathrm{s}=T_\alpha(H^*)$.~\cite{Maiolepsza:2015jj}
Henceforth, $T$ of gREM is represented by the temperature
determined from the peak of the enthalpy distribution for each replica from
Eq.~(\ref{eq:Teff}) (see also 
Fig.~\ref{fig:enthalpy_energy_distribution}(a) below).
In contrast, $T$ of tREM is represented by the designed temperature of each replica.

\begin{figure}[t]
\centering
\includegraphics[width=0.45\textwidth]{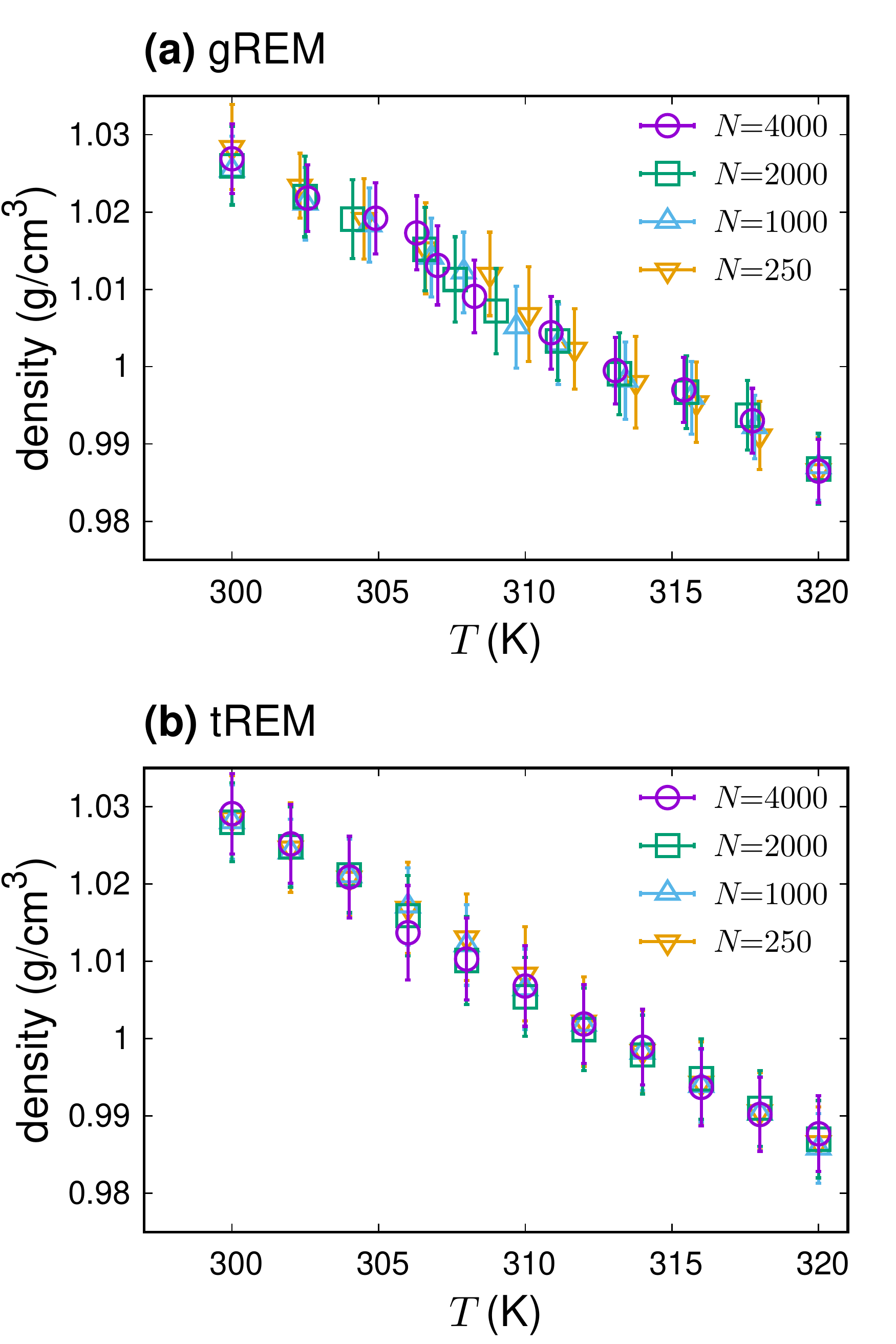}
\caption{
Temperature dependence of the average of density with different system
 sizes $N=250$, $1000$, $2000$, and $4000$, using the gREM (a) and tREM (b).
The bar associated with each point of the density corresponds to the standard
 deviation.
} 
\label{fig:density}
\end{figure}

\begin{figure}[t]
\centering
\includegraphics[width=0.45\textwidth]{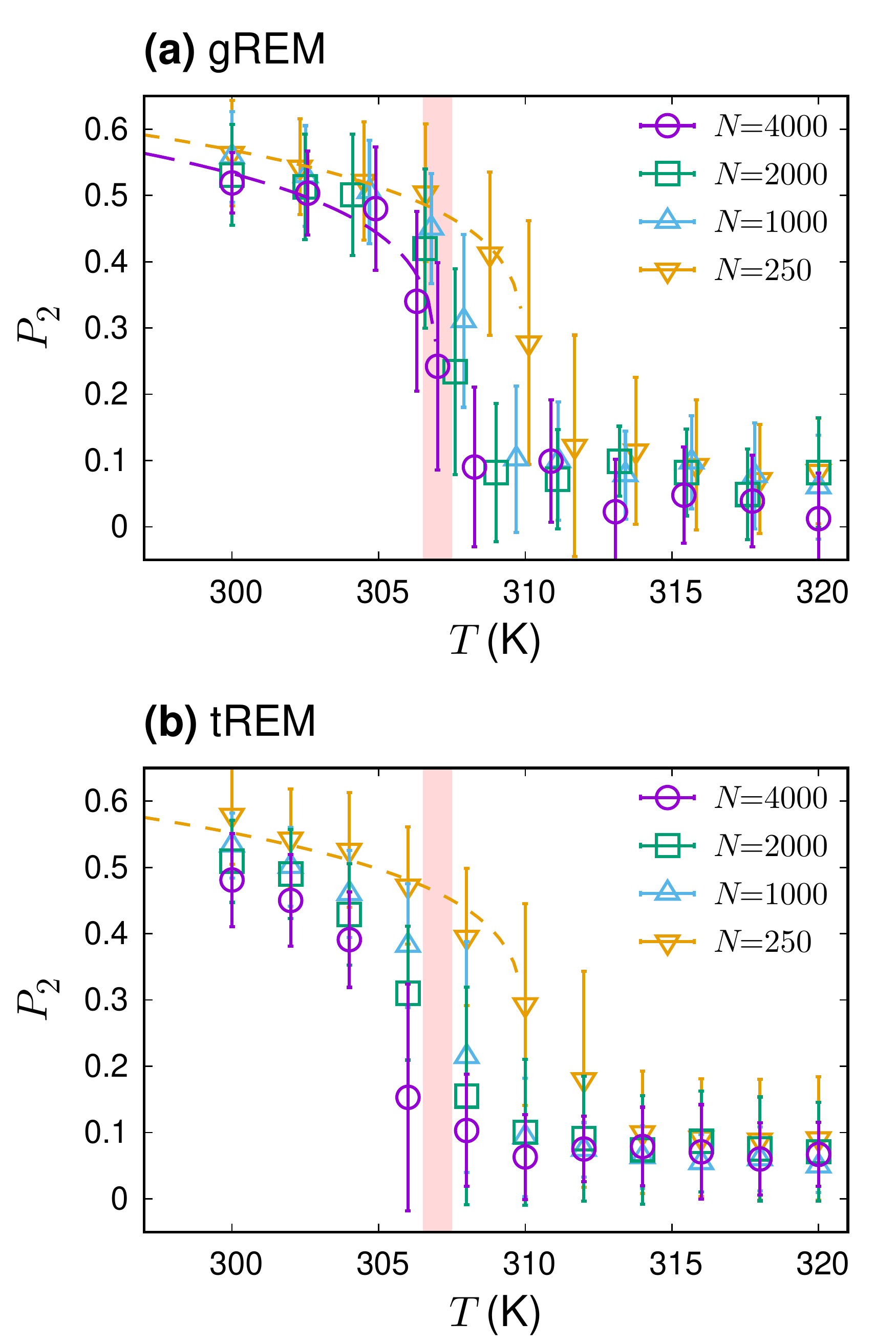}
\caption{Temperature dependence of the average of the order parameter $P_2$ with different system
 sizes $N=250$, $1000$, $2000$, and $4000$, using the gREM (a) and tREM (b).
The bar associated with each point of $P_2$ corresponds to the standard
 deviation.
In (a), the purple dashed curve represents the fitting result with
 Eq.~(\ref{eq:Haller}) for $N=4000$.
The vertical bar indicates the NI transition
 temperature, $T_\mathrm{NI}\approx 307$ K, determined from the fitting
 for results of $N=4000$.
In (a) and (b), the fitting results with
 Eq.~(\ref{eq:Haller}) for $N=250$ are described by 
the orange dashed curves.
The NI transition temperature is $T_\mathrm{NI}\approx 310$ K, which is
 consistent with the value reported in Ref.~\onlinecite{Tiberio:2009iq}.}
\label{fig:order_parameter}
\end{figure}

\begin{figure}[t]
\centering
\includegraphics[width=0.45\textwidth]{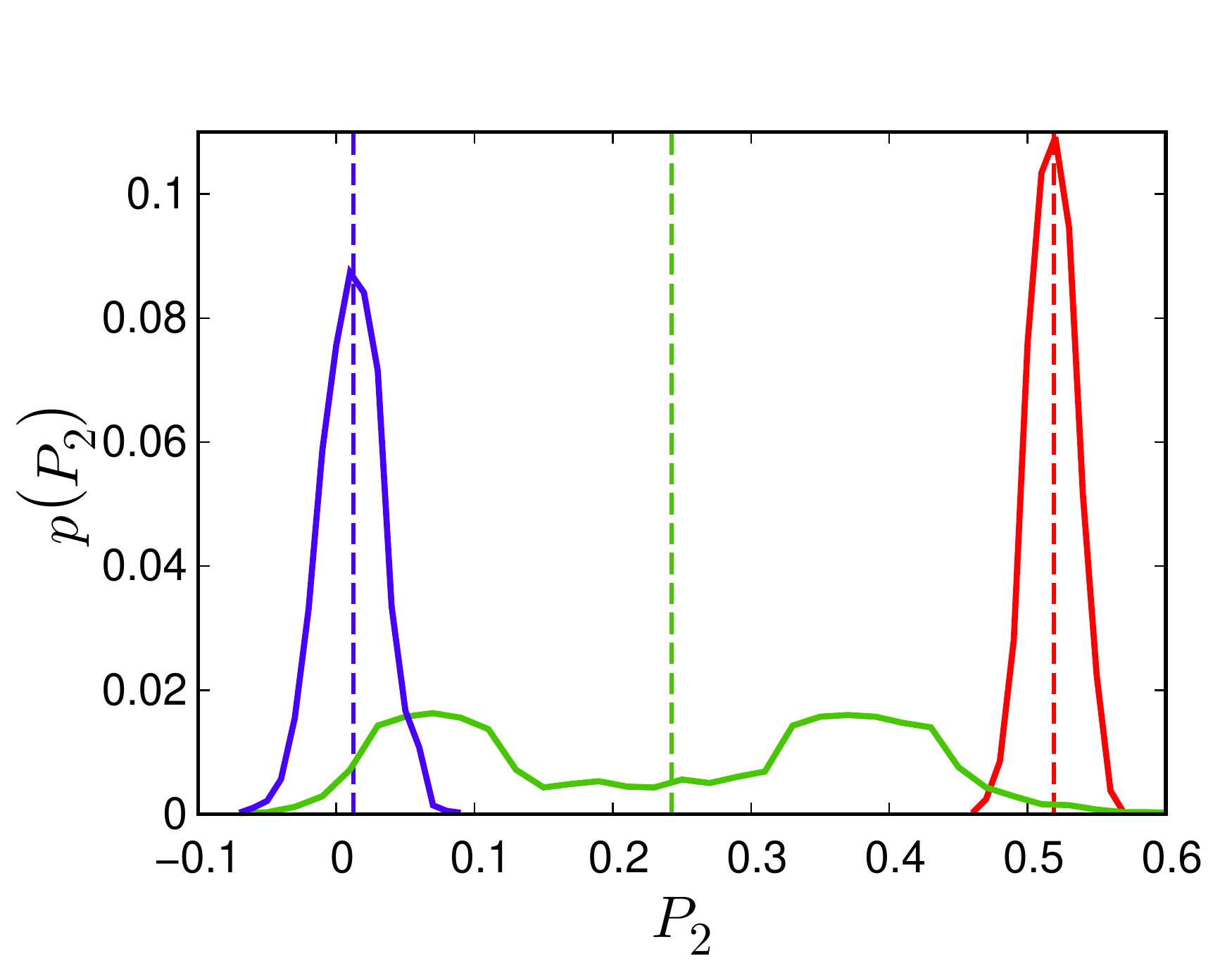}
\caption{
Probability distribution of the order parameter $P_2$ sampled in replicas 1 (blue),
 5 (green), and 11 (red) using the gREM with $N=4000$.
Their effective temperatures are 320, 307.0, and 300 K, respectively.
The vertical lines indicate the mean values, \textit{i.e.}, $P_2\approx 0.02$ (blue),
 0.26 (green),
 and 0.52 (red).
} 
\label{fig:order_parameter_distribution}
\end{figure}

\section{Results and discussion}
\label{sec:results}

First, we investigate the temperature dependence of the density.
Figure~\ref{fig:density} shows the results for gREM (a) and tREM (b) by
changing the number of molecules $N$.
The sudden drop of the density is evident at around 305-310 K,
particularly for the larger system size with $N=4000$.
The transition from nematic to isotropic
phases occurs in this temperature range, as demonstrated in
previous studies using the same UA
model.~\cite{Tiberio:2009iq, Sidky:2018hr,
Shi:2020cb, Sheavly:2020bn}

We next examine the orientational order parameter of the LC nematic
phase.
The order parameter $P_2$ is generally
expressed by the second-order Legendre polynomial as follows:
\begin{equation}
P_2 = \left\langle \frac{1}{N}\sum_{i=1}^N\left(\frac{3}{2} \cos^2 \theta_i - \frac{1}{2}\right)\right\rangle,
\end{equation}
where $\theta_i$ represents the angle between the director $\bm{n}$ and 
the unit vector of the long axis of molecular $i$,
$\bm{u}_i$.
Here, the director $\bm{n}$ is the unit vector representing the
preferred direction of the local volume.
The brackets denote the an ensemble averages.
The most widely used method to quantify $P_2$ from MD simulations is 
the diagonalization of the order parameter tensor
$\bm{Q}$.~\cite{Zannoni:1979tw, Allen:2017wd}
The expression of $\bm{Q}$ is given by 
\begin{equation}
\bm{Q} = \left\langle\frac{1}{N}\sum_{i=1}^N\left( \frac{3}{2}\bm{u}_i
					     \otimes \bm{u}_i -
					     \frac{1}{2}\bm{I}\right)\right\rangle,
\end{equation}
with the unit matrix $\bm{I}$.
The eigenvalues of the matrix $\bm{Q}$, $\lambda_-<\lambda_0<\lambda_+$,
guarantee $\lambda_-+\lambda_0+\lambda_+=0$ because $\bm{Q}$ is traceless.
The largest eigenvalue $\lambda_+$ provides the order parameter $P_2$,
and the corresponding eigenvector is the director $\bm{n}$. 
In practice, $P_2$ was calculated as $-2\lambda_0$, which was suggested
for a better estimation of $P_2$, particularly in the isotropic phase.~\cite{Eppenga:2006he}
Here, $\bm{u}_i$ in this work refers to the C$\equiv$N bond of the cyano group.
Note that the dependence of the choice of $\bm{u}_i$ on
$P_2$ was negligible if the inertia axis of the molecule is
used.~\cite{Tiberio:2009iq}

Figure~\ref{fig:order_parameter} shows the temperature dependence of the
order parameter $P_2$ from gREM (a) and tREM (b) simulations.
It is observed that 
the decrease in temperature $T$ leads to a decrease in $P_2$ from
0.5-0.6 to nearly zero for all the system sizes that were investigated.
The temperature dependence of $P_2$ 
exhibits a sharper drop across the NI transition for the larger system with $N=4000$.
However, $P_2$ for the smallest system size with $N=250$ gradually
decreases with increasing $T$.
In general, the discontinuity of the first-order phase transition between two
stable states are smeared out in finite systems.~\cite{Binder:1984iq, Binder:1997gn}
The $N$ dependence in Fig.~\ref{fig:order_parameter} is in
accordance with the well-known fact that unstable regions of the
phase transition become narrow as the system size decreases.

Here, we used the empirical equation,~\cite{Haller:1975eh, Chirtoc:2010kw}
\begin{equation}
P_2 = (1-P^\mathrm{iso}_2)\left(1-\frac{T}{T_\mathrm{NI}}\right)^{\beta} + P^\mathrm{iso}_2,
\label{eq:Haller}
\end{equation}
for fitting with the order parameter $P_2$.
We obtained the residual order parameter under the isotropic phase
$P^\mathrm{iso}_2\approx 0.12$, NI transition temperature
$T_\mathrm{NI}\approx 307$ K, and pseudo-critical exponent $\beta
\approx 0.20$ for the gREM result with $N=4000$.
The fitting result is shown in Fig.~\ref{fig:order_parameter}(a).
Note that $T_\mathrm{NI}\approx 307$ K is slightly lower than the
experimental value of 308 K.~\cite{Dalmolen:1984dv, Deschamps:2008cq}
In contrast, the fitting for tREM with $N=4000$
becomes uncertain due to the unsampled value of $P_2$ around the NI
transition temperature, as observed in Fig.~\ref{fig:order_parameter}(b).
The fittings of $P_2$ with $N=250$ using Eq.~(\ref{eq:Haller})
are also shown in Fig.~\ref{fig:order_parameter} (a) and (b).
For both gREM and tREM, the NI transition temperature is estimated as 
$T_\mathrm{NI}\approx 310$ K.
This value is in good agreement with the NI
temperature reported in Ref.~\onlinecite{Tiberio:2009iq}, where the number of
molecules was $N=250$.
Furthermore, 
$\beta \approx 0.23$ (gREM) and $\beta \approx 0.21$ (tREM) are also close to
the reported value of $0.226$.~\cite{Tiberio:2009iq}

The distribution of the orientational order parameter $P_2$ is
also important to examine the NI transition.~\cite{Tiberio:2009iq}
Figure~\ref{fig:order_parameter_distribution} shows the results from
replicas 1, 5, and 11 for the gREM with $N=4000$.
The corresponding temperatures $T_\alpha(H^*)$ of gREM are 300 K,
307.0 K, and 320 K, respectively.
The value of 307.0 K corresponds to $T_\mathrm{NI}$, which was determined
from the fitting results using
Eq.~(\ref{eq:Haller}).
The distributions of $P_2$ in replicas 1 and 10 are unimodal, whereas
two peaks at $P_2\approx 0.1$ and 0.4 are observed in replica 5,
resulting in the large fluctuation of $P_2$ at 307.0 K, as shown in
Fig.~\ref{fig:order_parameter}(a).
The bimodal distribution of the order parameter $P_2$ is consistent with
the free energy landscape predicted by the mean-field theory for
first-order phase transitions~\cite{deGennes:1995vq}. 
However, the theoretical exponent
$\beta=0.5$ is different from the simulation result.
In addition, it is interesting to examine
pretransitional behaviors in the isotropic phase because the NI
transition is relatively
weak first-order phase transition.
This regard can be characterized by the short-ranged orientational order from 
the orientational correlation functions $G_1(r)=\langle\delta
(r-r_{ij})(\bm{u}_i\cdot \bm{u}_j)\rangle_{ij}$ and $G_2(r)=\langle\delta
(r-r_{ij})[(3/2)(\bm{u}_i\cdot \bm{u}_j)^2-(1/2)]\rangle_{ij}$,~\cite{Tiberio:2009iq}
where $r_{ij}$ denotes the distance of center-of-mass between molecules
$i$ and $j$.
We also calculated $G_1(r)$ and $G_2(r)$ using the gREM with $N = 4000$ and obtained 
results consistent with those reported in
Ref.~\onlinecite{Tiberio:2009iq} (data not shown).

\begin{table*}[t]
\caption{\label{table:accpetance_ratio}
Acceptance ratio of the replica exchanges during the last 10 ns simulation 
 between two replicas $\alpha$ and
 $\alpha+1$ for the gREM and tREM with $N=4000$.
Note that * indicates 
that no exchanges were observed during 1000 trials.}
\begin{ruledtabular}
\begin{tabular}{ccccccccccc}
 replica index $\alpha$  & 1  & 2  &  3  & 4 & 5 & 6 & 7 & 8 & 9 & 10\\
\hline
 gREM (\%) & 14.8 & 13.6 & 12.8 & 12.9 & 12.5 & 14.6 & 17.4 & 18.1 & 18.3 & 18.9 \\
 tREM (\%) & 8.0 & 4.8 & 6.0 & 0.8 & * & 3.2 & 5.2 & 11.6 & 7.2 & 11.4 
\end{tabular}
\end{ruledtabular}
\end{table*}

\begin{figure}[t]
\centering
\includegraphics[width=0.45\textwidth]{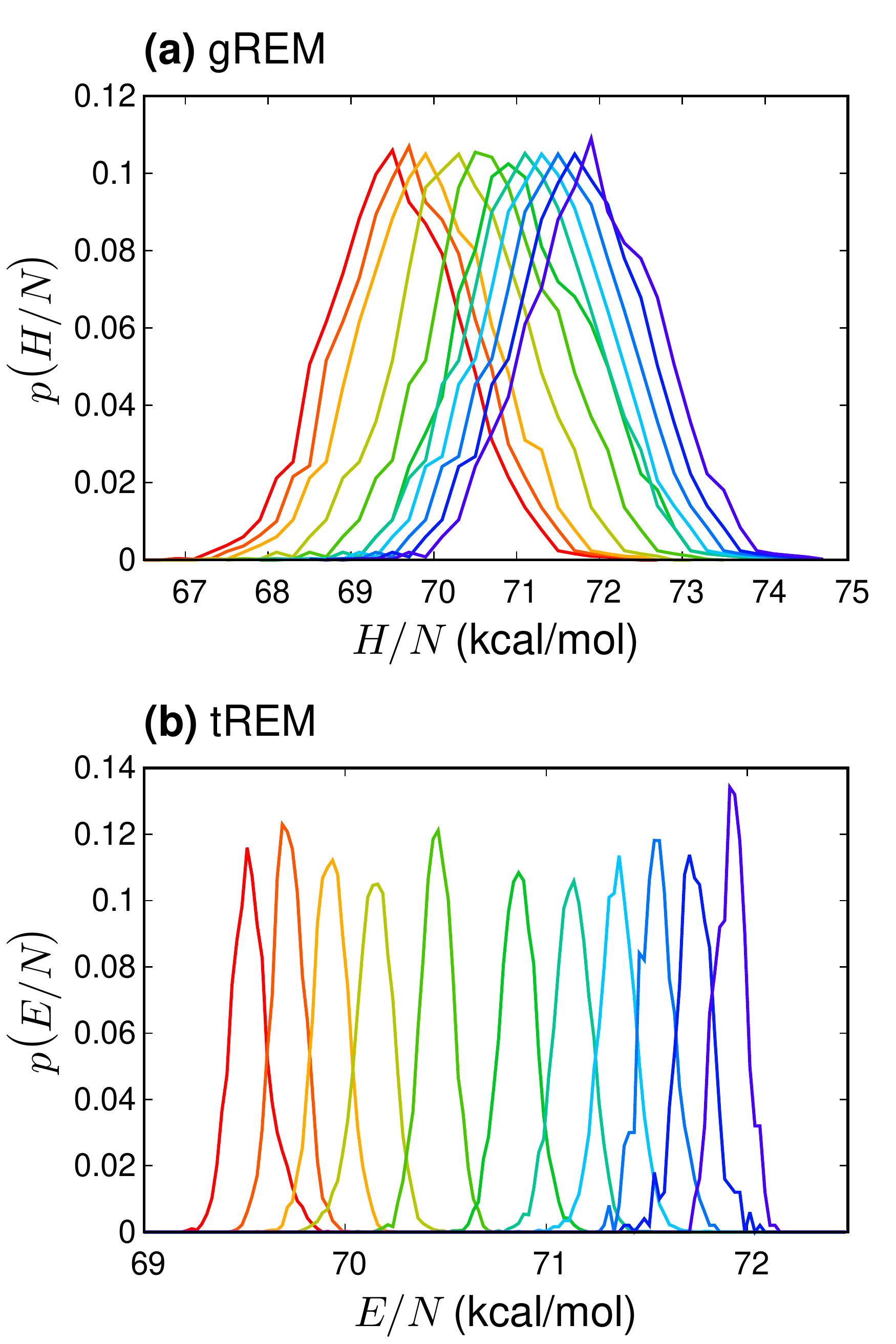}
\caption{
Probability distributions of enthalpy $H$ and potential energy $E$
for different replicas of the gREM (a) and tREM (b) with $N=4000$.
From left to right, the replica index $\alpha$ ranges from 1 to 11.
The enthalpy and potential energy are normalized by the number of molecules $N$.
} 
\label{fig:enthalpy_energy_distribution}
\end{figure}

\begin{figure}[t]
\centering
\includegraphics[width=0.45\textwidth]{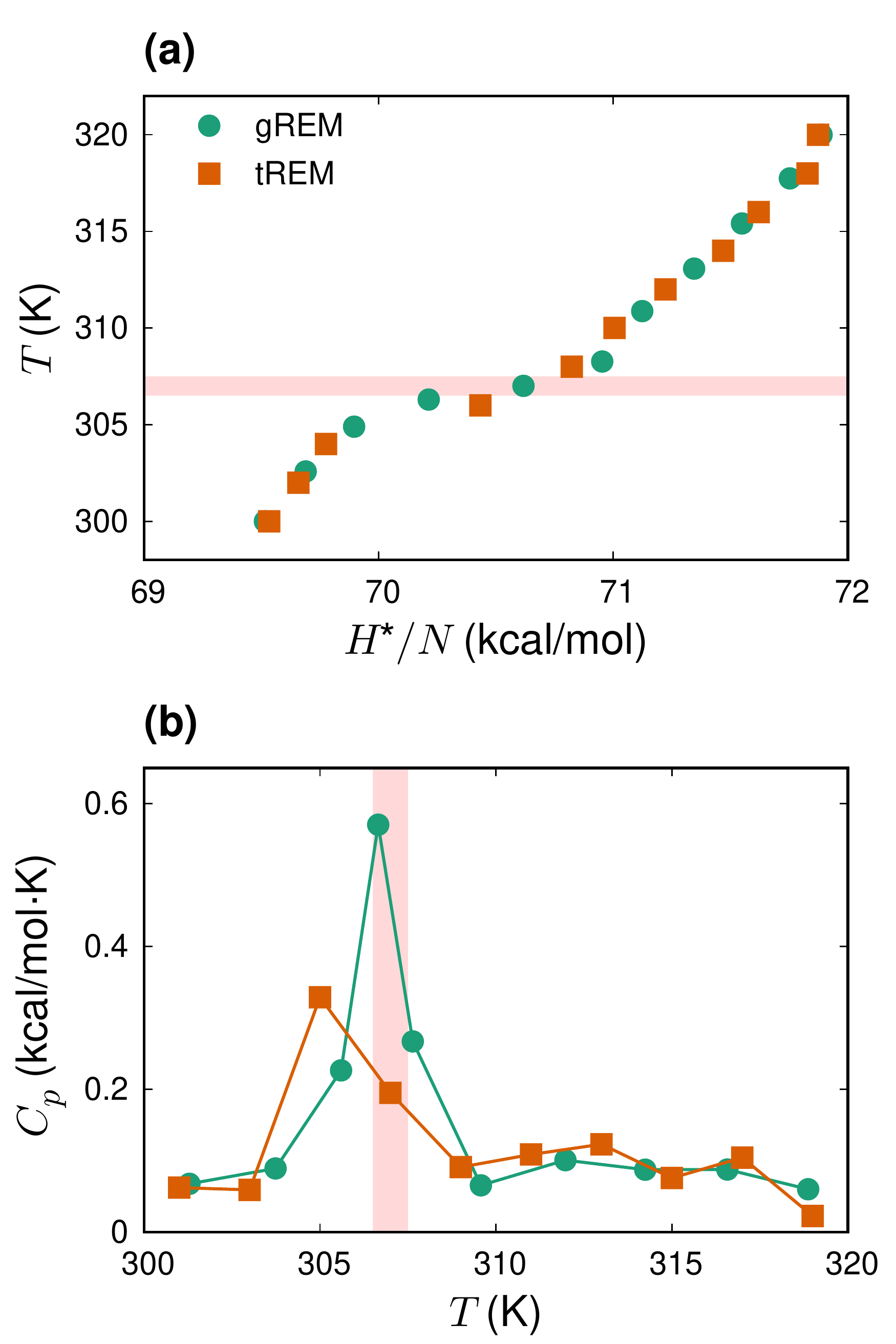}
\caption{
(a) The temperature $T$ as a function of the normalized enthalpy $H^*/N$
for gREM
 and tREM with $N=4000$.
$H^*$ represents the most probable enthalpy for each replica, resulting in
 the statistical temperature $T$ from the relation $T_\alpha(H^*)$ for gREM.
For tREM, $T$ represents the designed temperature of each replica.
From left to right, the replica index $\alpha$ ranges from 1 to 11.
(b) Heat capacity $C_p$ as a function of temperature $T$ for gREM
 and tREM with $N=4000$.
The horizontal bar in pink in (a) and the vertical bar in (b) indicate
 the NI transition temperature, $T_\mathrm{NI}\approx 307$ K,
 respectively.
} 
\label{fig:Teff}
\end{figure}

It should be noted that the difference between gREM and tREM becomes
significant for the system with $N=4000$ 
around $T_\mathrm{NI}$
(see Figs.~\ref{fig:density} and
~\ref{fig:order_parameter}).
Although the reduction of density and $P_2$ with temperature was also reproduced using
the tREM, 
the values of density and 
$P_2$ from the tREM and gREM were not in agreement in the transition region.
In particular,
the averaged $P_2$ at 306 K (replica 4) of the tREM in
Fig.~\ref{fig:order_parameter}(b) is close to the value
in the isotropic phase with a large error bar.
To elucidate the difference between the tREM
and gREM with $N=4000$, we quantified the acceptance ratios of the replica exchanges,
the results of which are summarized in Table~\ref{table:accpetance_ratio}.
The acceptance ratio in gREM is larger than that in tREM for any 
pair of replica exchanges.
The exchange ratio in gREM reduces between replicas 4 and 5, near $T_\mathrm{NI}\approx
307$ K.
Nevertheless, it exceeds 10\% and assures the efficiency of the replica-exchange method.
This is clear evidence that gREM is an effective simulation method, 
even near the phase transition temperature for larger system sizes.
By contrast, 
the replica exchanges between replicas 4 ($T=306$ K) and 5 ($T=308$ K) 
and between replicas 5 ($T=308$ K) and 6 ($T=310$ K) are significantly
restricted in tREM, which is an effect of the phase transition, as
anticipated beforehand.
We confirmed that the isobaric-isothermal REM relatively enhances the
replica exchange between neighboring replicas, but the acceptance ratio
between replicas 4 and 5 remains small at 0.2\%.
This is due to the fact that the volume change of the NI transition is 
small about 4\% from 300 K to 320 K (see Fig.~\ref{fig:density}).

The acceptance ratios between neighboring replicas are determined by the
degree of overlap in the enthalpy or potential energy distributions
in gREM or tREM, respectively.
Figure~\ref{fig:enthalpy_energy_distribution} shows the probability distribution 
of enthalpy in gREM (a) and potential energy in tREM (b) for each replica.
The transition region corresponds to replicas 4, 5, and 6. 
The overlap
of the distributions is significant in this region when gREM is employed.
However, the overlap becomes scarce when using tREM, 
leading to a significant reduction in replica-change events.
In principle, the acceptance ratios of 
replica exchanges in tREM can 
be improved by using more
replicas to increase the degree of overlap of the potential energy
distribution.
However, simulating the tREM with more replicas is impractical in terms of
computational costs, particularly for significantly larger system sizes.
In contrast, for the smallest system size with $N=250$, we
confirmed that replica exchanges of tREM remain even at the NI
transition region, showing 
behaviors in the density
and $P_s$ similar to those of gREM, as seen in Figs.~\ref{fig:density}
and ~\ref{fig:order_parameter}.

Finally, we examine the relationship between the temperature
$T$ and 
most probable enthalpy $H^*$, which is plotted in Fig.~\ref{fig:enthalpy_energy_distribution}(a).
Here, $T$ of gREM was determined from $T_\mathrm{s}=T_\alpha(H^*)$ using Eq.~(\ref{eq:Teff}) with 
the peak value $H^*$ of the enthalpy distribution for replica $\alpha$ (see
Fig.~\ref{fig:enthalpy_energy_distribution}(a)).
It has been demonstrated that $T$ becomes flat when crossing the
NI transition temperature ($T_\mathrm{NI}\approx 307$ K).
For comparison, the temperature as a function of $H^*$ from tREM simulations
with $N=4000$ is also shown in Fig.~\ref{fig:Teff}(a).
Although $T(H^*)$ of tREM also exhibits a curve similar to
that of gREM, the plotted points become sparse, particularly for the NI
transition region.
The temperature dependence of the heat capacity $C_p=\partial
(H^*/N)/\partial T$ 
with $N=4000$ can be further evaluated from a central difference
approximation for $H^*/N$
as a function of $T$, as plotted in Fig.~\ref{fig:Teff}(b).
The sharp peak associated with the NI phase transition was observed at
$T_\mathrm{NI}\approx 307$ K for gREM.
This peak is also consistent with the drop of the density at the NI 
transition with $N=4000$, as observed
in Fig.~\ref{fig:density}(a).
In contrast, the peak of $C_p$ becomes unclear for tREM, again
indicating the necessity of more replicas to increase the resolution around the NI
transition.
For smaller system sizes, the inflection of the $T-H^*/N$ curve is gradually
smeared out, causing smaller peaks of $C_p$ for
both gREM and tREM (data not shown).

\section{Conclusions}
\label{sec:conclusions}

In this paper, we report on MD simulation results combined with the gREM for
the NI phase transition of the 5CB model.
We demonstrated that the gREM can be applied to the NI phase transition of LC
systems.
It is effective in the sense that the acceptance ratios of the replica
exchange remain significant at the NI transition temperature even with
increasing $N$.
By contrast, the replica exchange of tREM became inefficient,
particularly for larger system sizes, although tREM simulated the results
similar to those with gREM for $N=250$.

The temperature dependence of the density and orientational order parameter $P_2$
was also examined.
In particular, $P_2$ in the transition region between the nematic and
isotropic phases can be sampled well using the gREM with the help of the
effective temperature.
A sharp drop in $P_2$ toward the NI temperature was observed, and this
was more prominent for larger system sizes.
In addition, a sharp peak of heat capacity $C_p$ was clearly observed around the NI
temperature.

An advantage of gREM 
is its availability for the smectic phase of 4-octyl-4'-cyanobiphenyl (8CB) using the
relevant UA model.~\cite{Palermo:2013fg}
Note that a recent work shows the free energy landscape of smectic-nematic phase
transition using machine learning technique.~\cite{Takahashi:2021bk}
It is interesting to examine these complex phase transition behaviors using gREM.
Moreover, 
it is important to investigate the applicability of gREM in 
all-atom MD simulations of LC systems, for example, self-assembling
helical structures~\cite{Yoshida:2018iq} and
nanochannels,~\cite{Ishii:2021hu} where phase transitions occur continuously over a wide temperature range.
Further studies focusing on these aspects are required.

\begin{acknowledgments}
The authors acknowledge Prof.~Go Watanabe of Kitasato University for helpful discussions.
K.K. is grateful to Prof.~Hajime Yoshino of Osaka University for valuable comments.
This work was supported by JSPS KAKENHI Grant Numbers:
JP19H05718~(H.W.), 
JP18H01188~(K.K.), JP19H01812~(K.K.), JP20H05221~(K.K.), and JP19H04206~(N.M.).
This work was also partially supported 
by the Fugaku Supercomputing Project (No.~\mbox{JPMXP1020200308}) and the Elements Strategy
Initiative for Catalysts and Batteries (No.~\mbox{JPMXP0112101003}) from the
Ministry of Education, Culture, Sports, Science, and Technology.
The numerical calculations were performed at Research Center of
Computational Science, Okazaki Research Facilities, National Institutes
of Natural Sciences, Japan.
\end{acknowledgments}

\section*{AUTHOR DECLARATIONS}

\section*{Conflicts of Interest}
The authors have no conflicts to disclose.

\section*{data availability}
The data that support the findings of this study are available from the
corresponding authors upon reasonable request.

%

\end{document}